\documentclass[aps,pre,preprint,groupedaddress,epsfig,superscriptaddress,showpacs]{revtex4-1}
\usepackage{graphicx}
\usepackage{bm}
\usepackage{CJKutf8}
\usepackage{caption} 
\usepackage{color}
\usepackage{ragged2e}
\renewcommand{\raggedright}{\leftskip=0pt \rightskip=0pt plus 0cm}

\begin{document}


\title{The coherent motions of thermal active Brownian particles}


\author{Cheng Yang}
\affiliation{School of Mathematics and Physics, Mianyang Teachers' College, Mianyang 621000, China}

\author{Ying Zeng}
\affiliation{School of Mathematics and Physics, Mianyang Teachers' College, Mianyang 621000, China}

\author{Shun Xu \footnote{E-Mail: xushun@sccas.cn
}}
\affiliation{Computer Network Information Center, Chinese Academy of Sciences, Beijing 100083, China}

\author{Xin Zhou\footnote{E-Mail: xzhou@ucas.ac.cn
}}
\affiliation{School of Physical Sciences, University of Chinese Academy of Sciences, Beijing 100049, China}


\date{\today}

\begin{abstract}
Active matter exhibits many intriguing non-equilibrium character, \emph{e.g.}, the active Brownian particles (ABP) without any attractive and aligned interactions can occur the mobility-induced phase transition to form some dense domains with both the structural ordering and dynamical coherence. Recently, the velocity correlation among the particles in the dense and ordered clusters was found in athermal ABP systems, however, seemed to disappear if including the thermal noises to describe microscopic ABPs, bringing some confusion about the generality of the consistence between structure and dynamics in ABPs. Here we demonstrate that the thermal noises imposing a large random term on the instantaneous velocity of ABPs hinder the observation of the (small) correlation in motions of ABPs. By averaging the instantaneous velocity (or equivalently, calculating the displacement) in various lag times, we show that the motions of thermal-fluctuated ABPs in the one order of magnitude smaller than the translational characteristic time are highly coherent and consistent spatially with the structural ordering of the ABPs.
\end{abstract}


\maketitle

\section{Introduction}
Active matter contains units that can consume exterior (or interior) energy to maintain their movements\citep{bechinger2016active,das2020introduction,marchetti2013hydrodynamics,
ramaswamy2010mechanics}. These units are different kinds of organisms, such as birds \citep{ballerini2008interaction}, cells\citep{angelini2010cell,garcia2015physics}, or some artificial particles such as self-propelled colloidal particles\citep{buttinoni2013dynamical,palacci2013living,theurkauff2012dynamic}. The active matter often presents  ordered structure and collective behaviors out of the description and expectation of the equilibrium statistical physics, such as, the ubiquitous patterned structure and collective motions in many physiological processes\citep{friedl2009collective,weijer2009collective,ilina2009mechanisms,shaw2009wound} and animal migrations\citep{buhl2006disorder}. A lot of attentions have been paid to investigating the underlying mechanism of the ordered structure and coherent dynamics, especially their relationships. 
In earlier studies, ones deemed that the anisotropic interactions between active units, such as the elongation of hard particles in shape\citep{ginelli2010large,peruani2006nonequilibrium,kudrolli2008swarming} or the explicit aligning interaction\citep{vicsek1995novel,ballerini2008interaction} was the key for the active particles moving coherently. 
More recently, spherical particles with only pure repulsive interactions were also found to occur the mobility induced phase separation (MIPS) to form clusters and patterns of dilute and dense steady states. The coherent motions were reported with high consistence to the dense steady states, by simulations \citep{wysocki2014cooperative,caprini2020spontaneous,caprini2020hidden,caprini2021spatial} and experiments\citep{deseigne2010collective,deseigne2012vibrated}. Among them, Caprini \emph{et. al.}\citep{caprini2020spontaneous} showed that the active Brownian particles (ABP), as a typical model system of active mater\citep{digregorio2018full,redner2013structure,fily2012athermal,bialke2013microscopic,speck2015dynamical,buttinoni2013dynamical}, underwent the MIPS and aligned the velocities of the clustered particles in the dense states. Since the thermal noise was neglected, their studied systems seemed to mimic the macroscopic ABP systems. 
Caporuso \emph{et al.}\citep{caporusso2020motility} simulated microscopic ABP systems by involving thermal diffusion, surprisedly, they did not observe any velocity correlations inside the dense clusters formed by the MIPS. The conflicting results brings an important open question, whether the correlation between the ordered steady structure and the collective dynamics is general in ABPs, which have aroused interesting discussions\citep{szamel2021long} and need to clarify. 
 
In this work, we focus on answering the question whether the motion coherence of high-dense structure-ordered active particles is broken by thermal noises. We simulate the two-dimensional (2D) pure repulsive spherical ABP systems, and we find that the thermal noise contributes directly a large random term to the instantaneous velocity of each ABP thus covers up the inherent coherence of ABP's motions. It is indeed the reason that the velocity correlation was not observed in relevant references.  We show that the time-averaging velocity of ABPs, \emph{i.e.} the displacement of ABPs over a lag time, can effectively depress the random part of the instantaneous velocity and preserve the inherent character of the ABP's motion. We find that there are collective-motion clusters in the dense states of ABPs formed by MIPS no matter if thermal noises exist or not, and these clusters with collective motions are highly related to the high hexagonal symmetry of the self-propulsion particles. Along the rim of the ordered clusters, self-propulsion forces mostly point inward and compress to sustain these clusters, wherein the particles move coherently to form some vortex-like or aligned domains. This work indicates that the correlation between the structured clusters and the dynamical domains is general in ABP systems regardless thermal noises. 

\section{Simulation}
The system contains $10000$ particles, in a square simulation box with length $L$ and periodic boundary conditions. These particles are governed by two overdamped Langevin equations\citep{redner2013structure},
\begin{equation}
\label{eq1}
\dot{\vec{r}}_i=-D \beta \vec{\nabla}_i U+D \beta F_p \vec{n}_i+\sqrt{2 D}\vec{\eta}_i,
\end{equation}

\begin{equation}
\label{eq2}
\dot{\theta}_i=\sqrt{2D_r}\eta^R_i.
\end{equation}
Here, the pure repulsive Weeks-Chandler-Anderson potential $U(r_i)=\sum_{j\ne i} 4\epsilon [(\frac{\sigma}{r_{ij}})^{12}-(\frac{\sigma}{r_{ij}})^{6}]+\epsilon$ is truncated at $\sigma_d=2^{\frac{1}{6}}\sigma$. The parameter $\sigma$ is the diameter of particles. For simplicity, we set $\epsilon=k_BT$ where $k_B$ is the Boltzmann constant. $D$ and $D_r$ are translational and rotational diffusion constants, respectively. They are related by $D_r=\frac{3D}{\sigma^2}$, and $\beta=\frac{1}{k_B T}$. $F_p$ is the magnitude of the self-propulsion and $\vec{n}_i=(cos \theta_i, sin \theta_i)$ denotes its orientation. $\vec{\eta}_i$ and $\eta^R_i$ are zero-mean Gaussian white noise with $\langle\eta(t)\eta(t')\rangle=\delta(t-t')$. In all simulations, we choose $\sigma$,\ $\tau=\frac{\sigma^2}{D}$ and \ $k_BT$ as the units of length,time and energy, respectively. The area packing fraction $\phi=\pi \sigma^2 N/(4L^2)$ varies from $0.40$ to $0.55$ and P{\'e}clet number $P_e=\frac{F_p \sigma}{k_B T}$ is set to $100$. Each trajectory was run for $100 \tau$ with time step $10^{-5}\tau$.  
    
\section{Results}
\begin{figure}[ht]
\includegraphics[width=0.8\textwidth]{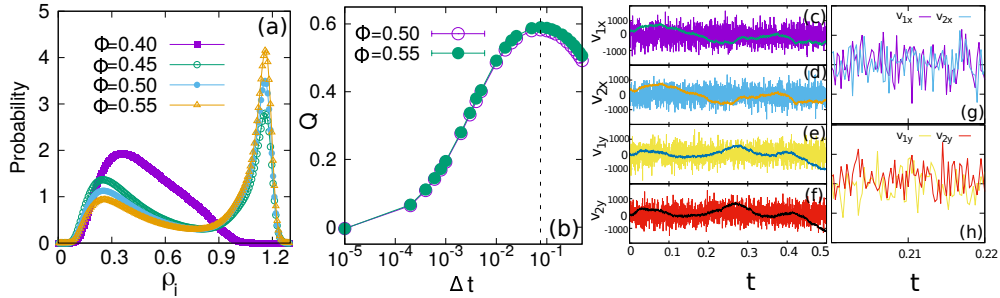}
\caption{(a) Probability distribution functions of the local density at different packing fractions. (b) In high-density phase, order parameter $Q$ varies with $\Delta t$ and its maximum occurs at $0.075\tau$ indicated by the black dotted line. In (c)$\sim $(f) we compare velocities of two neighbor particles in high-density phase at $\phi=0.50$: (c) $V_{1x}$ (purple) and $100 \times \overline{V_{1x}}$ (green), (d) $V_{2x}$ (azure) and $100 \times \overline{V_{2x}}$ (orange), (e) $V_{1y}$ (yellow) and $100 \times \overline{V_{1y}}$ (blue), (f) $V_{2y}$ (red) and $100 \times \overline{V_{2y}}$ (black). The mean velocities were multiplied by $100$ for their small values and the lag time $\Delta t = 0.075 \tau$. (g) Comparison of $V_{1x}$ and $V_{2x}$. (h) Comparison of $V_{1y}$ and $V_{2y}$.}
\label{Fig.1}
\end{figure}

We first investigate probability distribution functions (PDFs) of the local density at different area packing fractions. The local density is calculated through $\rho_i=\frac{\pi \sigma^2}{4 v_i}$, where the local volume $v_i$ is obtained with the Voronoi tessellation algorithm\citep{rycroft2009voro++}. As shown in Fig.\ref{Fig.1}(a), these distributions are bimodal when the packing fraction is bigger than $0.4$, consistent with the prior result\citep{redner2013structure}. The low (high)-$\rho$ peak corresponds to the low (high)-density phase. In the following of this article, we will consider the phase separation case only.

Before exploring the velocity correlation, we define the mean velocity at time $t$ as $\overline{\vec{V}_i}(t, \Delta t)=\frac{\vec{r}_i(t+\Delta t)-\vec{r}_i(t)}{\Delta t}$, here $\Delta t$ is the lag time. If $\Delta t \rightarrow 0$, the mean velocity degenerates into an instantaneous one. The velocity correlation can be described by the order parameter $Q= \langle q_i \rangle_h$, where $q_i=1-2 \sum_{ij} \frac{d_{ij}}{N_i \pi}$ \citep{caprini2020spontaneous}. $i$ is the central particle and $j$ is one of its nearest neighbors (found by the software Voro++\citep{rycroft2009voro++}). $d_{ij}$ is the angle between the velocities of particle $i$ and $j$. $N_i$ denotes the number of neighbors. $\langle \dots \rangle_h$ represents taking the average over all particles in the high-density phase. In Fig.\ref{Fig.1}(b), we show that there are no correlations among instantaneous velocities ($Q=0$ as $\Delta t \rightarrow 0$). Then the correlation increases gradually over the lag time. The maximum of $Q$ is reached at $\Delta t = 0.075 \tau$ and we will discuss this maximum in detail later.

To have an intuitive impression of velocity correlations, we choose two close neighbor particles in the high-density phase and compare their velocities in both x and y directions. In Fig.\ref{Fig.1}(c) [or (d)], we show the instantaneous velocity $V_{1x}$ (or $V_{2x}$) and the corresponding mean velocity, $100 \times \overline{V_{1x}}$ ( or $100 \times \overline{V_{2x}}$).  In Fig.\ref{Fig.1}(e) [or (f)], we show the instantaneous velocity $V_{1y}$ (or $V_{2y}$) and the corresponding mean velocity, $100 \times \overline{V_{1y}}$ ( or $100 \times \overline{V_{2y}}$).  As shown in Fig.\ref{Fig.1}(c)$\sim$(f), mean velocities of these two particles are correlated obviously. The comparison of $V_{1x}$ and $V_{2x}$ (or $V_{1y}$ and $V_{2y}$) is shown in Fig.\ref{Fig.1}(g) [or (h)]. It is hard to observe correlations between instantaneous velocities directly. As the mean velocities are much smaller than the instantaneous ones, the thermal fluctuations will contribute mostly and the inherent velocities will be ignored when we calculate the correlation of instantaneous velocities.

\begin{figure}[ht]
\includegraphics[width=0.8\textwidth]{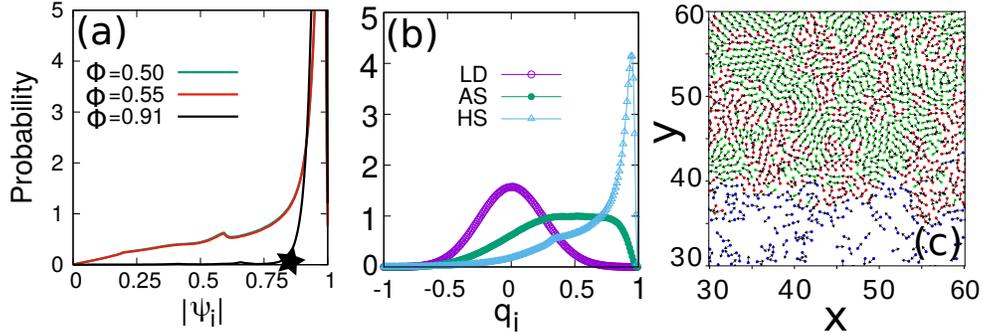}
\caption{(a) Probability distribution functions of $|\psi_i|$ in high-density phase at different packing fractions. The black line is obtained from simulation at $P_e=0$ and $\phi=0.91$. The black star indicates the criterion distinguishing the hexagonal symmetrical and asymmetrical particles. (b) Probability distribution functions of $q_i$ for different kinds of particles at $\phi=0.5$ and $\Delta t=0.075 \tau$. (c) A steady snapshot at $\phi=0.5$. Black arrows indicate orientations of mean velocities with $\Delta t = 0.075 \tau$. Colors encode hexagonal symmetrical (green), asymmetrical (red) and low-dense (blue) particles, respectively.}
\label{Fig.2}
\end{figure}  

We proceed to discuss the relationship between velocity correlation domains and ordered clusters of ABPs. The local structure can be depicted by the bond-orientational order parameter\citep{redner2013structure}. It is defined as $\psi_i=\frac{1}{N_i}\sum_j e^{i 6 \theta_{ij}}$, where $N_i$ is the number of nearest neighbors of particle $i$. $\theta_{ij}$ is the angle between the $i$-$j$ bond and x-axis, where $j$ represents the neighbor. The larger the $|\psi_i|$, the higher the hexagonal symmetry of local structures. As shown in Fig.\ref{Fig.2}(a), PDF of $|\psi_i|$ in the high-density phase exhibits a tall peak near one and a long tail extending to zero. The peak represents hexagonal symmetrical local structures, while the tail indicates asymmetrical ones. To distinguish them clearly, we perform an equilibrium simulation with $P_e=0.0$ and $\phi=0.91$ (hard-sphere close-packing value\citep{digregorio2018full,redner2013structure}). This PDF has a tall peak near one only, implying most particles are in the hexagonal symmetrical state. We choose $0.8$ as the criterion to distinguish the hexagonal symmetrical and asymmetrical local structures. Combining with PDFs in Fig.\ref{Fig.1}(a), we obtain three types of particles: hexagonal symmetrical (HA) particles (particles with hexagonal symmetrical local structures in the high-density phase), asymmetrical (AS) particles (particles with asymmetrical local structures in the high-density phase), and low-dense (LD) particles (particles in the low-density phase). 

\begin{figure}[ht]
\includegraphics[width=0.6\textwidth]{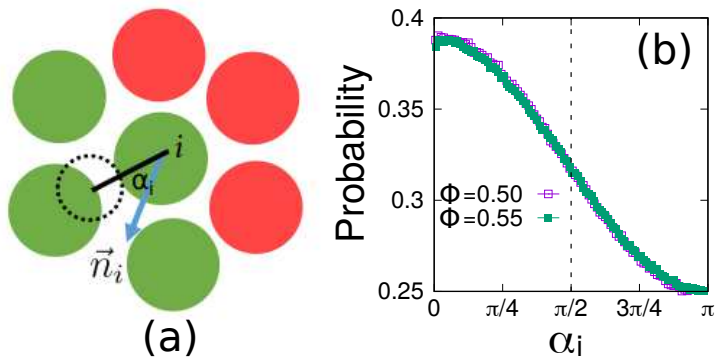}
\caption{(a) The schematic illustrating of the order parameter $\alpha_i$, where $i$ is the central particle. Green and red particles represent hexagonal symmetrical and asymmetrical particles, respectively. The dotted circle denotes the center of mass (CM) of symmetrical neighbors. $\alpha_i$ is defined as the angle between $i$-CM bond and $\vec{n}_i$. (b) PDFs of $\alpha_i$ at different area packing fractions.}
\label{Fig.3}
\end{figure}  

We next investigate the velocity correlation of different types of particles. The distributions of $q_i$ are shown in Fig.\ref{Fig.2}(b). Hexagonal symmetrical particles have the strongest velocity correlation, followed by asymmetrical particles, and low-dense particles have the weakest correlation. A snapshot of different types of particles is shown in Fig.\ref{Fig.2}(c). Black arrows indicate orientations of mean velocities. In the high-density phase, there are many velocity correlation domains. Colors encode different types of particles and hexagonal symmetrical particles form many ordered clusters in the high-density phase, surrounded by asymmetrical particles. The velocity correlation domains and the ordered clusters are highly coincident spatially. 

To understand the origin of velocity correlation domains, we introduce an order parameter $\alpha_i$ in Fig.\ref{Fig.3}(a). Particle $i$ is hexagonal symmetrical and has at least one hexagonal symmetrical neighbor and one asymmetrical neighbor, i.e., it is at the edge of the ordered cluster. The dotted circle represents the center of mass (CM) of all the hexagonal symmetrical neighbors. The angle between $\vec{n}_i$ and the $i$-CM bond is $\alpha_i$. PDFs of $\alpha_i$ are shown in Fig.\ref{Fig.3}(b). Interestingly, the probability of acute angle is higher than that of obtuse angle, implying the self-propulsions of the edge particles point to the inside of the cluster mostly. The ordered cluster are compacted tightly by the edge particles. As a result, particles in the cluster move coherently giving rise to the velocity correlation domains.

\begin{figure}[ht]
\includegraphics[width=0.4\textwidth]{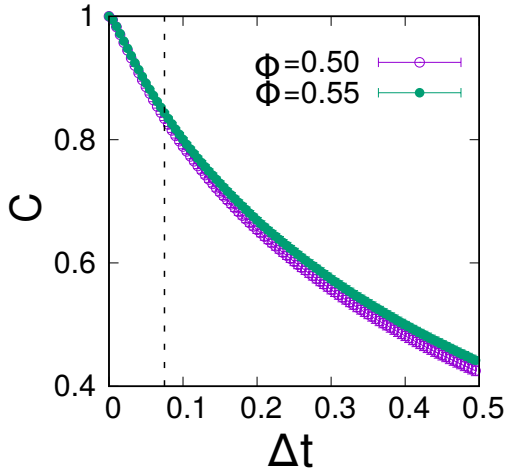}
\caption{The order parameter $C$ changes over the lag time at different packing fractions. The dotted line indicates $\Delta t=0.075\tau$, around which $C \sim 5/6$, implying one neighbor has been replaced.}
\label{Fig.4}
\end{figure}  

We next discuss the dynamics of hexagonal symmetrical particles. Let $N_i(t)$ denote the number of neighbors of particle $i$ at time $t$, and $M_i(t+\Delta t)$ is the remaining number of neighbors that haven't been replaced at $t+\Delta t$. $C(\Delta t)=\langle \frac{M_i(t+\Delta t)}{N_i(t)}\rangle_{HS}$ is the ratio, here $\langle \dots \rangle_{HS}$ represents taking the average over all hexagonal symmetrical particles. In Fig.\ref{Fig.4}, we show $C$ varies over $\Delta t$. At both packing fractions, $C$ decreases rapidly. The value of $C$ is around $5/6$ at $0.075 \tau$, representing that about one neighbor has been replaced. Less than this time (no neighbors have been replaced), the ordered cluster behaves like a solid, while beyond this time the cluster begins to deform. It is worth noting that the velocity correlation $Q$ also reaches its maximum at $0.075 \tau$ [see Fig.\ref{Fig.1}(b)]. The value of $Q$ can be affected by two factors: (i) the longer the lag time, the smaller the influence of thermal fluctuations; (ii) if the lag time is too long, the cluster begins to deform, which will reduce the velocity correlation. Under the joint action of these two factors, $Q$ reaches the maximum when the ordered cluster begins to deform. 

\section{Conclusion}
In conclusion, the velocity correlations of two-dimensional overdamped ABPs are observed. The ordered clusters formed by particles with hexagonal symmetrical local structures are the key to understanding the origin of the velocity correlation domains. At the boundary of these ordered clusters, the self-propulsions of particles point inward mostly, which compresses to sustain the cluster. Thus particles in these clusters move coherently leading to the velocity correlation domains.  

\section{Acknowledgments}
This work was supported by the Initial Scientific Research Fund of Mianyang Teachers' College (Grant No. QD2020A03) and GHFUND A (Grant No. 202107011618).

\bibliographystyle{apsrev4-1}
\bibliography{ref.bib} 


\end{document}